\documentclass[12pt]{article}
\usepackage{amsmath}
\usepackage[utf8]{inputenc}
\usepackage[a4paper,left=2cm,right=2cm,top=2cm,bottom=2cm]{geometry}
\def\Lag{{\mathcal L}{}}
\def\Sag{{\mathcal S}{}}
\def\Mag{{\mathcal M}{}}
\def\Bag{{\mathcal B}{}}
\def\tref{{}^{0}h}
\def\gref{{}^{0}g}
\def\Lw{{\mathcal L}{}}

\def\omw{{\stackrel{\bullet}{\omega}}{}}
\usepackage{setspace}

\def\Kw{{\stackrel{\bullet}{K}}{}}
\usepackage[colorlinks=true,urlcolor=blue]{hyperref}
\usepackage{etex}
\usepackage{amsmath,amsfonts,amsbsy,amssymb,amscd,latexsym}
\usepackage{times}
\usepackage{verbatim}
\usepackage[sort&compress,numbers]{natbib}
\usepackage[affil-it]{authblk}

\begin{document}
\title{ \bf  Einstein gravity from the Einstein action: Counterterms and covariance}
\author{Martin Kr\v{s}\v{s}\'ak\thanks{Electronic address: \texttt{martin.krssak@gmail.com, martin.krssak@fmph.uniba.sk}}}
\affil{Department of Theoretical Physics, Faculty of Mathematics, Physics and Informatics, Comenius University in Bratislava, 84248, Slovak Republic}
\affil{Department of Astronomy, School of Physical Sciences, University of	Science and Technology of China,  96 Jinzhai Road, Hefei, Anhui 230026, China}
\date{\today}
\maketitle

\begin{abstract}
The field equations of general relativity can be  derived from the Einstein action, which is quadratic in connection coefficients, rather than the standard action  involving the Gibbons-Hawking-York term and counterterm. We show that it is possible to construct a new counterterm directly for the Einstein action, which removes divergences and naturally introduces a flat reference spacetime. The  total action is then covariant under simultaneous transformation of both the spacetime and reference tetrads, and argue that this is analogous to the  Gibbons-Hawking action. We then explore different  perspectives arising naturally from different uses of the reference tetrad, and explore implications of viewing gravity as fundamentally described in terms of non-covariant connection coefficients.
\end{abstract}

\section{Introduction\label{secintro}}
The standard approach to general relativity \cite{Landau:1982dva, Weinberg, Misner:1974qy}, which mostly follows  the original Einstein's path \cite{Einstein:1915by,Einstein:1915ca}, starts with studying the motion of a free-falling particle and identifying gravity with the spacetime geometry. The non-tensorial Christoffel symbols are recognized to play  the role of the total inertial and gravitational forces acting on the particle.  However, when we formulate the field equations for the geometry itself, we usually invoke  the principle of covariance and demand that the field equations are tensorial. We are then lead to consider the Riemann tensor, removing all of the nontensorial coordinate-dependent behaviour of Christoffel symbols,  and we are practically uniquely led to Einstein field equations.

An interesting situation occurs if we try to  derive these field equations from the variational principle. As is well-known, it was Hilbert who almost simultaneously  with Einstein derived the field equations from the action principle using what is nowadays known as the  (Einstein-)Hilbert action \cite{Hilbert:1915tx},  given by the scalar curvature
\begin{equation}\label{ehaction}
	\Sag_\text{H}=
	\frac{1}{2\kappa} \int_{\mathcal{M}} \sqrt{-g} \, R,
\end{equation}
where $\kappa=8\pi$ in the natural units.
The remarkable property of this action is that  the scalar curvature contains second derivatives of the metric, and hence one would naively expect fourth-order field equations. However, as it turns out,  the higher derivatives  form a non-dynamical total derivative term and we are left with second-order field equations.

It was recognized only in the 1970s by York \cite{York:1972sj} and Gibbons and Hawking \cite{Gibbons:1976ue} that, due to the total derivative term,   we can  derive the field equations only if we require variations of both the metric and their normal derivatives to vanish at the boundary. This generally represents a consistency problem for the variational problem, and can be solved by adding the  so-called Gibbons-Hawking-York (GHY) boundary term \cite{York:1972sj,Gibbons:1976ue}  
\begin{equation}\label{ghyapp}
	\Sag_\text{GHY}=
	\frac{1}{\kappa} \int_ {\partial \mathcal{M}} \sqrt{-\gamma} \mathcal{K},
\end{equation}
where $\mathcal{K}$ is the extrinsic curvature of the boundary   and $\gamma_{\mu\nu}$ is the induced metric on the boundary. The GHY term \eqref{ghyapp} is chosen in a such way that its variation cancels out the variation of the total derivative term  in \eqref{ehaction}, and then the Einstein field equations can be derived consistently, i.e., assuming only  variations of the metric  to vanish at the boundary \cite{Poisson:2009pwt}.

The motivation of Gibbons and Hawking to consider \eqref{ghyapp} was not just to vary the action and obtain the field equations, but to find a finite Euclidean gravitational action solutions for studying  black hole thermodynamics using path integrals, i.e. to evaluate the action on-shell \cite{Gibbons:1976ue,Hawking:1978jz}. Therefore, adding \eqref{ghyapp} turned out to be crucial since  \eqref{ehaction} vanishes in vacuum and hence the GHY term dominates the action.

However,  including  the GHY term renders the action  divergent in the general case, requiring regularization by adding a counterterm. The total gravitational action  then takes   the  form
\begin{equation}\label{lagtot1}
	\Sag_\text{grav}=\Sag_\text{H}+\Sag_\text{GHY} +\Sag_\text{counter},
\end{equation}
and we face the challenge of finding an appropriate counterterm to eliminate divergences. We are primarily interested in removing IR divergences coming from the limit $r\rightarrow \infty$, since the Euclidean time becomes periodic. In  asymptotically flat spacetimes, the original proposal by Gibbons and Hawking \cite{Gibbons:1976ue}, known nowadays as the \textit{background subtraction}, was to add a counterterm given by the GHY term for the reference or background metric $\gref_{\mu\nu}$, which is a flat Minkowski metric in which the spacetime metric is isometrically embedded.  The total gravitational action \eqref{lagtot1} then depends on both the spacetime and reference metrics  and can be written as
\cite{Gibbons:1976ue,Hawking:1995fd}
\begin{equation}\label{lagtot11}
	\Sag_\text{grav}(g,\gref)=\Sag_\text{H}(g)+\Sag_\text{GHY}(g) +\Sag_\text{counter}(\gref).
\end{equation}
The well-known problem is that the isometric embedding of a metric in Minkowski spacetime does not have to  exist, what lead to development of new regularization methods, including holographic renormalization, where the counterterm is a local covariant function of the intrinsic geometry of the boundary \cite{Balasubramanian:1999re,Mann:2005yr}, and the recent  counterterm for null boundaries \cite{Parattu:2015gga}.

Besides the challenge of determining the appropriate counterterm in specific situations, the  presence of the counterterm can be seen as ``somewhat awkward" \cite{Hawking:1979ig} and naturally leads to questions that has
 likely perplexed anyone encountering the full gravitational action \eqref{lagtot11} for the first time: How come  that in a fully covariant theory that excludes any background structures, the divergences are removed using the reference/background metric? Moreover,
what  is actually being subtracted from the gravitational action?

\section{Einstein actions \label{sec1916}}
While Einstein originally did not derive the field equations from the action principle, he did introduce his own action in October 1916 \cite{Einstein:1916cd}, when he proposed the  Lagrangian
\begin{equation}\label{LagEin}
	\Lag_\text{E}= \frac{1}{2\kappa}\sqrt{-g}g^{\mu\nu}(
	\Gamma^\rho{}_{\sigma\mu}\Gamma^\sigma{}_{\rho\nu}-
	\Gamma^\rho{}_{\mu\nu}\Gamma^\sigma{}_{\rho\sigma}
	),
\end{equation}
which can be obtained from the Hilbert Lagrangian by separating the contribution of the second derivative terms into a total derivative term \cite{Landau:1982dva,Dirac1975,Padmanabhan:2010zzb}
\begin{equation}\label{decomp}
	\Lag_\text{H}=\Lag_\text{E}+\Lag_\text{tot}.
\end{equation}
Varying this Lagrangian with respect to the metric, we do obtain the  vacuum field equations in their potential form \cite{Einstein:1916cd}, where we essentially separate the  terms containing second derivatives of the metric and the terms quadratic in first derivatives of the metric \cite{Freud1939,Bohmer:2017zid}. The quadratic terms will be given by    \cite{Einstein:1916cd}
\begin{equation}\label{Einsteinpseudo}
	t^\mu{}_\nu=\frac{1}{2}\left[\Lag_\text{E} \delta^\mu{}_\nu 
	-
	\frac{\partial \Lag_\text{E}}{\partial (\partial_\mu g^{\rho\sigma})} \partial_\nu g^{\rho\sigma}
	\right],
\end{equation}
known at the Einstein energy-momentum pseudotensor, from which the gravitational energy-momentum can be defined. Additionally, other pseudotensors, such as the symmetric pseudotensor proposed by Landau and Lifshitz \cite{Landau:1982dva}, or Weinberg \cite{Weinberg}, can also be introduced. 

The key characteristic shared by all these pseudotensors is that they are not tensors;  they are expressions quadratic in Christoffel symbols, and hence can always be made to vanish at a point using the Riemann normal coordinates.  In order to find meaningful predictions for the total energy-momentum, these pseudotensors need to be  evaluated in  well-behaved coordinate systems, which usually asymptotically approach the inertial coordinate system \cite{Landau:1982dva}. Despite their non-tensorial dependence on the coordinates, these pseudotensors do provide  correct answer for the gravitational energy-momentum that agrees with the Hamiltonian method \cite{Chang:1998wj,Chen:2018geu}.

The tetrad version of \eqref{LagEin} was rather accidentally discovered by Einstein in  the attempt to unify gravity and electromagnetism in late 1920s.  To this end, we consider a set of orthonormal vectors $h_a$ called the tetrad at each point of spacetime, with $h^a=h^a{}_\mu dx^\mu$ being its inverse and having components $h^a{}_\mu$ in some coordinate basis.
The tetrads form a non-coordinate basis and hence  the components of the spacetime metric can be written as
\begin{equation}\label{mettet}
g_{\mu\nu}=\eta_{ab}h^a_{\ \mu} h^b_{\ \nu},
\end{equation}
where $\eta_{ab}=\text{diag}(-1,1,1,1)$, allowing us to use the tetrads instead of the metric as a fundamental variable. The  tetrads  $h^a{}$  are independent of  coordinates and change under transformation of a  non-coordinate basis as
\begin{equation}\label{tettransf}
h^a\rightarrow \Lambda^a{}_b h^b,	
\end{equation} 
where $\Lambda^a{}_{b}$ is  a local Lorentz transformation to ensure preservation of orthonormality.

We can define the coefficients of anholonomy $f^c{}_{a b} = h_a{}^{\mu} h_b{}^{\nu} (\partial_\nu
h^c{}_{\mu} - \partial_\mu h^c{}_{\nu} )$, and follow Einstein\footnote{ Actually Einstein  presented this in a very different way in terms of the  torsion tensor that was supposed to represent both gravity and electromagnetism in the unified theory \cite{Sauer:2004hj}. However, as explained in the recent paper \cite{Krssak:2024xeh}, Einstein's ``torsion tensor" is actually just the coefficient of anholonomy and hence he effectively did what is presented here. We return to the teleparallel framework and torsion in Section~\ref{secequiv}.}  to  search for a Lagrangian quadratic in the coefficients of anholonomy that yields symmetric field equations. 
Einstein  found  such a Lagrangian in 1929 \cite{Einstein1929a}
\begin{equation}\label{LagE2}
	\Lw_\text{E}= -\frac{h}{2 \kappa} \left(\frac{1}{4} f^\rho{}_{\mu\nu}f_\rho{}^{\mu\nu}+\frac{1}{2} f^\rho{}_{\mu\nu} f^{\nu\mu}{}_{ \rho} -f^{\nu\mu}{}_{\nu} f^{\nu}{}_{\mu\nu}\right).
\end{equation}
While this Lagrangian looks naively very different from both the Hilbert \eqref{ehaction} and Einstein \eqref{LagEin} Lagrangians,  recalling that the Ricci rotation coefficients are related to the coefficients of anholonomy through \cite{Krssak:2024xeh}
\begin{equation}\label{lccon}
	\omega^a{}_{bc}=\frac{1}{2} \Big[f_b{}^a{}_c + f_c{}^a{}_b - f^a{}_{bc}\Big],
\end{equation}
we can recast it as
\begin{equation}\label{BulkLag}
	\Lag_\text{E}=	 \frac{h}{2\kappa} 
	\left(\omega^{a}_{\,\,\,ca}\omega^{bc}{}_{b}
	-\omega^{a}{}_{cb} 
	\omega^{bc}{}_{a} 
	\right).
\end{equation}
From here it should be obvious that  \eqref{BulkLag} is a tetrad version of \eqref{LagEin}, and hence Einstein in 1929 indeed ``just" rediscovered his earlier  1916 Lagrangian \eqref{LagEin} in tetrad formalism \cite{Krssak:2024xeh}. 

The alternative approach is to start with the tetrad Hilbert action, i.e. a scalar curvature in terms of the Ricci rotation coefficients \eqref{lccon}, and separate the total derivative term along the lines of \eqref{decomp}, resulting in \eqref{BulkLag}. This was first done by M\o ller in 1961 \cite{Moller1961}, and hence \eqref{BulkLag} is often referred as the M\o ller Lagrangian. It is then possible to define an analogue of the Einstein pseudotensor known as the  M\o ller  complex, which is a coordinate-tensor but a pseudotensor with respect to \eqref{tettransf}, from which  the total energy-momentum can be defined  as well \cite{Pellegrini1963,Moller1966,Virbhadra:1990zr}.

\section{Counterterms for the Einstein action}
We can now use the fact that the Hilbert action can be decomposed as \eqref{decomp} in both the metric and tetrad formulation, and   rewrite the total gravitational action \eqref{lagtot1} as
\begin{equation}\label{lagtot2}
	\Sag_\text{grav}=\Sag_\text{E}+\Sag_\text{tot}+\Sag_\text{GHY} +\Sag_\text{counter}.
\end{equation}
We can then observe that the dynamics is fully contained in the Einstein term $\Sag_\text{E}$, from which the field equations are derived, and which is divergent, similar  to \eqref{lagtot11} before adding the counterterm.
This brings us to the idea that the total derivative and GHY terms are not necessary in the total gravitational action \eqref{lagtot2}, and are rather just remnants of the mathematical formulation where the Hilbert action \eqref{ehaction} is the starting point. We would like to  argue here that it is possible to find a counterterm directly for the Einstein action, i.e. write the total gravitational action as
\begin{equation}\label{lagtot3}
	\Sag_\text{grav}=\Sag_\text{E} +\Sag_\text{counter},
\end{equation}
where the counterterm  is capable  to not only regularize  the Einstein action, but also ensure that the total gravitational action \eqref{lagtot3} is as covariant as the Gibbons-Hawking  gravitational action \eqref{lagtot11}.

To construct the counterterm, we  choose  to  do so for the tetrad \eqref{BulkLag} Einstein action
, but the construction for the metric Einstein action \eqref{LagEin} is analogous. We use the fact that the Einstein action \eqref{BulkLag}  under a change of the non-coordinate basis \eqref{tettransf},
transforms as\footnote{This can be obtained from a rather lengthy but straightforward calculation, or seen directly from the equivalent teleparallel formulation discussed in Section~\ref{secequiv} and using the results of \cite{Krssak:2015lba}.} 
\begin{equation}\label{equivLag}
	\Lag_\text{E}\rightarrow \Lag_\text{E}+ \frac{1}{\kappa} \partial_\mu
\left[ 	h  h_a{}^\nu h_d{}^{\mu} \eta^{bd}\Lambda^a_{\ c} \partial_\nu (\Lambda^{-1})^c_{\ b}\right],
\end{equation}
and  consider a \textit{reference tetrad}  representing a general tetrad in Minkowski spacetime
\begin{equation}\label{refftet}
\tref^a{}= \Lambda^a{}_b \tilde{\tref}{}^b,	
\end{equation} 
where $\tilde{\tref}{}^b$ is the  tetrad with components $\tilde{\tref}{}^b_{\ \mu}=\text{diag}(1,1,1,1)$  in the Cartesian coordinate system. 
Since the Riemannian spin connection\footnote{The Riemannian spin connection in the general spacetime is defined from the Ricci rotation coefficients \eqref{lccon} as $
		\omega^a{}_{b\mu}=\omega^a{}_{bc}h^c{}_\mu$}
vanishes for $\tilde{\tref}{}^b$, we find that  for the reference tetrad \eqref{refftet}   it can be written as
\begin{equation}\label{spin0def}
\omega^a{}_{b\nu} (\tref^a{})= 	\Lambda^a_{\ c} \partial_\nu (\Lambda^{-1})^c_{\ b},
\end{equation}
which follows from transformation properties of the spin connection \cite{AP}, and is precisely the term appearing in \eqref{equivLag}. 

We then choose the counterterm as   the total derivative term in \eqref{equivLag} expressed in terms of the reference tetrad \eqref{refftet}, i.e.
\begin{equation}\label{counterlag2}
	\Lag_\text{counter}(h^a,\tref^a)= \frac{1}{\kappa} \partial_\mu
	\Bag^\mu(h^a,\tref^a)=
	\frac{1}{\kappa} \partial_\mu\left[
h  h_a{}^\nu h_d{}^{\mu} \eta^{bd} \omega^a{}_{b\nu} (\tref^a{})
	\right],
\end{equation}
and  the full gravitational action  takes the form
\begin{equation}\label{lagtot4}
	{\Sag}_\text{grav}(h^a,\tref^a)=\Sag_\text{E}(h^a) +\Sag_\text{counter}(h^a,\tref^a),
\end{equation}
where $h^a$ and $\tref^{a}$ are the spacetime and reference tetrads, respectively.

Based on the analogy with the Gibbons-Hawking action \eqref{lagtot11}, we require that the total action \eqref{lagtot4} vanishes for the Minkowski spacetime, what is achieved if we identify the full tetrad with the reference tetrad,  i.e. $h^a=\tref^a$.  In asymptotically flat spacetimes, we require that the spacetime tetrad reduces to the reference tetrad in the limit $r\rightarrow\infty$, same as in the case of Gibbons-Hawking action \eqref{lagtot11}.
 

Let us illustrate this on the example of the Schwarzschild solution. We take the diagonal tetrad
\begin{equation}\label{tetdiag}
h^a{}_\mu=\text{diag}(f,f^{-1},r,r\sin\theta), \qquad f^2=1-\frac{2M}{r},
\end{equation}
for which the Einstein action \eqref{BulkLag} is
\begin{equation}\label{actschwein}
\Sag_\text{E}=\frac{1}{\kappa}\int_\Mag  \sin\theta=\frac{1}{2}\int d t  \left. r\, \right|_{r_A}^{r_B}, 
\end{equation}
where $r_A$ and $r_B$ are the bounds of integration. In the case of asymptotically flat spacetime, $r_B\rightarrow\infty$ and the action diverges regardless of  whether we choose $r_A$ to be at the origin or the horizon.

We can then consider the reference tetrad $\tref^a{}_\mu=\text{diag}(1,1,r,r\sin\theta)$, and calculate the counterterm as
\begin{equation}\label{key}
\Sag_\text{counter}=\frac{1}{\kappa}\int_\Mag \frac{2M+(f-2)r}{fr}\sin\theta.
\end{equation}
Adding it to \eqref{actschwein} and integrating the spatial part, we obtain
\begin{equation}\label{schwaction}
	\Sag_\text{grav}=\Sag_\text{E}+\Sag_\text{counter}=\int d t \left. r(1-f)\right|_{r_A}^{r_B}. 
\end{equation}
However, when performing the integration leading to \eqref{schwaction}, we must be careful about the singularity and discontinuity of the Lagrangian at the horizon. We need to ensure that we integrate only over the manifold patches where the function is smooth. Therefore, we can consider the action either for the exterior or 	the interior solution of the black hole. The exterior solution is relevant in the Euclidean case, which covers only the exterior region of a black hole and time becomes periodic with a period $\beta=8\pi M$. We then choose $r_A=2M$ and $r_B\rightarrow \infty$, and  find the Euclidean action $S_\text{grav}=-i \beta M$, which is exactly twice the value of the original result by Gibbons and Hawking \cite{Gibbons:1976ue} using \eqref{lagtot11}\footnote{ Note that the same factor 2 was noticed for the metric Einstein action \eqref{LagEin}, or rather its teleparallel formulation discussed in Section~\ref{secequiv},  in \cite{BeltranJimenez:2018vdo}. Moreover, it was suggested to the use of the so-called  canonical frames could resolve the problem of the factor of 2  \cite{BeltranJimenez:2019bnx}. For an alternative resolution, see our upcoming paper 
\cite{KrssakStano}.
}.

The interior solution is relevant for the recently proposed  complexity=action duality, where it corresponds to the Wheeler-de Witt (WdW) patch at late times \cite{Brown:2015bva}. In this case, we are primarily interested not in the action itself but rather   its time derivative known as the action growth. Choosing  $r_A=0$ and $r_B=2M$, we find the action growth at the WdW patch to be $d\Sag_\text{grav}/d t=2M$ \cite{Krssak:2023nrw}, which exactly agrees with the standard GHY method  for the interior region of a black hole \cite{Brown:2015bva}. Moreover, this agreement for the interior solutions can be found not only in the Schwarzschild case, but also when the electric charge and negative cosmological constant are included \cite{Krssak:2023nrw}.

\section{Four faces of the  gravitational action \label{secequiv}}
The total gravitational action can be then viewed from four  different perspectives. Besides the standard Gibbons-Hawking  action \eqref{lagtot11}, the Einstein action with the appropriate counterterm \eqref{lagtot4} is  a  viable alternative.
Moreover, two further perspectives on the gravitational action can be found by using the reference tetrad in different ways.

The first option is to effectively eliminate   $\tref^a$  and view the theory within the special class of  tetrads. We start with $\{h^a,\tref^a\}$  and consider some local Lorentz transformation  $\Lambda^a{}_b$ that transforms $\tref^a$ to the Cartesian tetrad $\tilde{\tref}^a$, what turns the counterterm into zero and the reference tetrad will effectively disappear from the theory. We then obtain a special class of tetrads for which the Einstein action  \eqref{BulkLag} will give the correct regularized action without a  counterterm.  These are the so-called  \textit{proper tetrads} known from teleparallel gravity   \cite{Lucas:2009nq,Krssak:2015rqa, Krssak:2024xeh}.

The advantage of this perspective is that the existence of proper tetrads explains  why the M\o ller complex evaluated in a special class of tetrads gives the correct finite prediction for the energy-momentum: the proper tetrads are the tetrads in which the counterterm \eqref{counterlag2} is trivial, and  the Einstein action \eqref{BulkLag} is regular and hence the M\o ller complex derived from it gives the correct regularized conserved charges.  An analogous argument  in the metric formalism for the original Einstein's 1916 action \eqref{LagEin} explains why the Einstein \eqref{Einsteinpseudo} and other pseudotensors have to be evaluated within preferred coordinate systems to achieve sensible finite predictions: these preferred coordinate systems are the coordinate basis in which the metric analogue of \eqref{counterlag2} is trivial. 

The last option is to  move away from the realm of Riemannian geometry and recast \eqref{lagtot4} as the action of teleparallel gravity \cite{AP,Krssak:2018ywd,Krssak:2024xeh}. We just have to use the fact  that   \eqref{spin0def}  can be identified as the teleparallel connection $\omw^a{}_{b\mu}$ utilized in teleparallel gravity \cite{AP,Obukhov:2002tm,Krssak:2018ywd}, and  use it instead of $\tref^a$.
The total action \eqref{lagtot4} can be then written fully in terms of the Riemannian $\omega^a{}_{b\mu}$  and teleparalel  $\omw^a{}_{b\mu}$ connections. These two connections are related through the Ricci theorem, which states that the difference between two such connections is necessarily proportional to the contortion tensor
\begin{equation}\label{key}
	\Kw^a{}_{b\mu}=\omw^a{}_{b\mu}-\omega^a{}_{b\mu}.
\end{equation}
We can then  recast the total Lagrangian \eqref{lagtot4} as \cite{AP,Krssak:2024xeh}
\begin{eqnarray}
	\Lw_\text{TG}(h^a{}_\mu,\omw^a{}_{b\mu})=
	- \frac{h}{2 \kappa} \left[\Kw^{abc}\Kw_{cba}-\Kw^{ac}{}_{a}\Kw^b{}_{cb}\right]. \label{lagtegr}
\end{eqnarray}

The equivalence can also be seen  from the fact that the resulting total action in the Schwarzschild case \eqref{schwaction} is equivalent to the results obtained within the teleparallel framework using either working with proper tetrads or calculating the spin connection \cite{Krssak:2023nrw,Krssak:2015rqa,Krssak:2015lba}. For further discussion and alternative viewpoints, see  \cite{Oshita:2017nhn,BeltranJimenez:2018vdo,Fiorini:2023axr}.

The advantage of the teleparallel form of the action  is that \eqref{lagtegr} is expressed in terms of the tensorial quantities and is manifestly covariant under the simultaneous transformation of the tetrad and teleparallel connection. Therefore, we can view teleparallel geometry as a geometric framework where the Einstein action \eqref{BulkLag} is naturally covariantized \cite{Krssak:2024xeh}, albeit in a sense that we discuss in the following section.

\section{Covariance of the gravitational action \label{seccov}}

While both the Einstein term \eqref{BulkLag} and the counterterm \eqref{counterlag2} are not covariant under a change of tetrad \eqref{tettransf}, their combination in the total action  \eqref{lagtot4} is   covariant under a simultaneous transformation of both the tetrad and reference tetrad
\begin{equation}\label{transfsim}
	\{h^a,\tref^a\}	\rightarrow
	\{\Lambda^a{}_b\, h^b,\Lambda^a{}_b\, \tref^b\}.
\end{equation}

Note that this is different from  covariance under a change of $h^a$ alone, and is analogous to the situation in the teleparallel framework, where the action \eqref{lagtegr} is covariant under simultaneous transformation of both the tetrad and spin connection. This is often viewed as an analogue  of the St\"uckelberg trick \cite{BeltranJimenez:2017tkd,BeltranJimenez:2018vdo}, and is recently a matter of discussion whether this  covariance should be taken seriously or considered artificial \cite{Maluf:2018coz,Golovnev:2023yla,Krssak:2024xeh}.

We have constructed the regularized Einstein action \eqref{lagtot4} in analogy with the construction of the full gravitational action \eqref{lagtot11}, and hence it should be obvious that it is as natural/artificial as  the full  gravitational action of general relativity \eqref{lagtot11}. Therefore,  the  covariance of \eqref{lagtot4} or \eqref{lagtegr} should not be discussed in comparison with the Hilbert term alone, but rather with respect to the total action, including all the boundary terms \eqref{lagtot11}.

While the Hilbert action is covariant with respect to transformations of the metric/tetrad alone, the covariance of the boundary terms is a rather subtle issue. 
The reason is that  the GHY term is covariant with respect to transformation of the boundary metric, which is determined through $\gamma^{\mu\nu}=g^{\mu\nu} -n^\mu n^\nu$, where $n^\mu$ is a normal vector to  the boundary.  Therefore, the GHY term  is covariant  only under simultaneous transformation of both $g^{\mu\nu}$ and $n^\mu$ in the bulk spacetime. Indeed, the extrinsic curvature scalar $\mathcal{K}=\nabla_\mu n^\mu$ is a scalar  only if we transform both the bulk metric and normal vector simultaneously. 

There are then only two options to keep the total gravitational action \eqref{lagtot11}  covariant. The first  is to transform  both the spacetime and reference metrics simultaneously, what is a direct analogue  of \eqref{transfsim}. In this case, a non-tensorial change  of the GHY term under a change of the spacetime metric alone is  counterbalanced by a non-covariant change of the GHY term for the reference metric.  The other viable option is to keep the reference metric fixed, and transform  both the spacetime metric and the normal vector simultaneously\footnote{Although this option is rather unconventional since the reference metric is not in the same coordinate system as the spacetime metric. It is  then not clear how the reference metric can be properly matched to the full metric.}, what ensures  the covariance of the GHY term and, hence,  the total action \eqref{lagtot11}. In both cases,  the total gravitational action \eqref{lagtot11} is not covariant under transformation of the spacetime metric alone, requiring transformation of  some additional structure: either the reference metric or the normal vector.

\section{Discussion and conclusions}
The connection coefficients play the key role in  general relativity where they represent accelerations measured by the observers. However, due to the  principle of covariance, they are used only through the curvature tensor and covariant derivatives derived from them, effectively eliminating all of their non-tensorial behavior from the theory.  While this is certainly  a reasonable demand for the field equations, it appears to be incompatible with the action principle. If we demand a finite action with a well-posed variational principle, we are forced to supplement the Hilbert term by the GHY term and an appropriate counterterm \eqref{lagtot11}. However, these terms turn out to depend not only on the bulk metric  but also on  foliations and/or the reference metric, which must  be transformed simultaneously with the metric to render the total action covariant.

We have argued here in favor of reappraisal of the Einstein  actions, \eqref{LagEin} and \eqref{BulkLag}, which are quadratic in connection coefficients. These actions do contain all the dynamics necessary to derive the field equations, but transform by a surface term under a change of basis that cause divergences. It is then  possible to use this feature and define a counterterm that removes divergences and makes the total action \eqref{lagtot4} covariant under simultaneous transformation of both the spacetime and reference tetrads \eqref{transfsim}. We have  argued that this analogous to covariance of the Gibbons-Hawking action \eqref{lagtot11}. 

Note that the idea of reference/background spacetimes predates the Gibbons-Hawking background subtraction and is a recurring topic, traceable back to Rosen's 1940 covariantization of pseudotensors using a reference metric \cite{Rosen:1940zza}. The reference metric was subsequently used in the positivity proof of gravitational energy \cite{Nester:1981bjx}, addressing the Komar anomaly \cite{Katz1985}, and calculating the angular momentum of the Universe \cite{BicakKatz}. Recent works have also considered different variational principles for the gravitational action \cite{Harada:2020ikm, Feng:2021lfa}.

In this paper, we have primarily focused on the tetrad formulation and the relevance of the reference/background spacetime on the value of the action. While the teleparallel formulation naturally introduces the reference spacetime through the teleparallel spin connection, we have demonstrated here how this construction can be made quite naturally within Riemannian geometry based on the analogy with the standard Gibbons-Hawking term \eqref{lagtot11}.
As a result, we were able to show analogies among the Gibbons-Hawking action \eqref{lagtot11}, the Einstein action with the counterterm \eqref{lagtot4}, and the teleparallel action \eqref{lagtegr}, as well as provide an explanation for why the original Einstein action \eqref{BulkLag} had to be evaluated in some privileged tetrad. All these different viewpoints emerged naturally depending on how we used the reference tetrad $\tref^a$.

Our goal was to  highlight the fundamental role of connection coefficients in general relativity and to illustrate that the  covariance of the field equations does not necessarily justify the common assertion that only covariant objects are  meaningful.  The  covariant field equations  can be derived from the non-covariant Einstein action, and this action, when supplemented by  an appropriate counterterm \eqref{lagtot4}, is as meaningful (and as covariant) as the Gibbons-Hawking action \eqref{lagtot11}.

In particular, this is relevant to the discussion of teleparallel gravity, where the covariance of the teleparallel action is often compared to that of the Einstein-Hilbert action.
Here, we have argued that the covariance of the teleparallel action should instead be compared to the full gravitational action of general relativity \eqref{lagtot11}, which is also covariant under the simultaneous transformation of both the full and reference spacetimes. Our construction of a counterterm for the Einstein action \eqref{lagtot4} can be then viewed as an interesting step 
towards understanding  the relation between the covariance in teleparallel gravity and general relativity.

This has important implications for the discussion about the equivalence of teleparallel gravity and general relativity. Teleparallel gravity is usually viewed as being only dynamically equivalent to general relativity, meaning it leads to the same field equations, but their actions differ by a total derivative term. Once again, this difference is there only if we are making the comparison with the Einstein-Hilbert action, which is not a full action. If instead, we compare the teleparallel action with the full Gibbons-Hawking action \eqref{lagtot11}, we find that the values of both actions--aside from an anomalous factor of 2--coincide. If the values of these actions coincide and they are both covariant in a similar fashion, this suggests that teleparallel gravity is not only dynamically, but a fully equivalent theory to general relativity.

Our approach naturally explains the necessity to subtract some reference structure from the gravitational action. If the gravitational action is covariant only under simultaneous transformation \eqref{transfsim}, it is clear that a reference tetrad or metric must be introduced. This also provides a rather straightforward explanation of what is being subtracted from the gravitational action using the reference spacetime. By acknowledging that the fundamental action is constructed from the connection coefficients, we are merely removing the effects associated with the non-tensorial character of the connection coefficients. One can then argue that these are the effects related to the choice of an observer, i.e., inertial effects.

While our discussion here was primarily focused on the tetrad formalism, a similar construction could be made by restricting ourselves to a class of coordinate frames, which would be related to previous results \cite{Nester:1981bjx,Katz1985,BicakKatz,Harada:2020ikm, Feng:2021lfa}. In this case, the teleparallel framework for such an action would be symmetric teleparallelism \cite{Nester:1998mp,BeltranJimenez:2017tkd}. A more interesting generalization  would be to relax the condition of orthonormality of frames \eqref{mettet}, i.e. to consider general $GL(4)$ frames instead of $SO(1,3)$ frames. If we would then try to find a geometric framework for such an action, we would find that this would lead to the general teleparallel geometries, i.e. those that contain both torsion and non-metricity, which were introduced recently \cite{Adak:2023ymc,Gomes:2023hyk}.

\section{Acknowledgements}
We would like thank Nathalie Deruelle and Justin C. Feng for bringing to our attention interesting references \cite{Nester:1981bjx,Katz1985,BicakKatz,Harada:2020ikm,Feng:2021lfa}.
This work was funded through  SASPRO2 project \textit{AGE of Gravity: Alternative Geometries of Gravity}, which has received funding from the European Union's Horizon 2020 research and innovation programme under the Marie Skłodowska-Curie grant agreement No. 945478.  

\bibliography{references}
\bibliographystyle{Style}
\end{document}